\documentclass[11pt,amstex,aps,amsfonts,prsty,preprint]{article} 
\usepackage[dvips]{graphicx}
\usepackage{amssymb}
\usepackage{amsmath}
\usepackage{a4wide}
\usepackage{citesort}
\newcommand{\be}{\begin{equation}}
\newcommand{\ee}{\end{equation}}

\begin{document}

\begin{titlepage}

\begin{center}

{\Large \bf Biased 
Tracer Diffusion in Hard-Core Lattice Gases: Some Notes on the 
Validity of the Einstein Relation.}

\vspace{0.1in}

{\Large  G.Oshanin$^{1}$, O.B\'enichou$^2$, S.F.Burlatsky$^3$ and M.Moreau$^{1}$}

\vspace{0.1in}

{$^{1}$ \sl Laboratoire de Physique Th{\'e}orique des Liquides, \\
Universit{\'e} Paris 6, 4 Place Jussieu, 75252 Paris, France}

\vspace{0.1in}

{\sl $^2$ Laboratoire de Physique de la Mati{\`e}re Condens{\'e}e, \\
Coll{\`e}ge de France, 11 Place M.Berthelot, 75252 Paris Cedex 05, France
}
\vspace{0.1in}

{$^{3}$ \sl United Technologies Research Center,\\
United Technologies Corporation,\\ 
411 Silver Lane, 129-21 East Hartford, CT 06108, USA
}

\vspace{0.1in}

\begin{abstract}
In this presentation we overview some recent results
on biased tracer diffusion in lattice gases. We consider
both models in which the gas particles
density is explicitly 
conserved and situations in which the lattice gas particles 
undergo continuous exchanges with
a reservoir, which case is appropriate, e.g., 
to adsorbed monolayers in contact 
with the vapor phase. For all these 
models we determine,  in some cases exactly and in other ones - 
using a certain decoupling approximation, 
the mean displacement of a tracer particle (TP) driven 
by a constant external force in a dynamical 
background formed by the
lattice gas particles whose transition rates are symmetric. 
Evaluating the TP mean displacement explicitly
we are able
to define
the TP mobility, which allows us to demonstrate that 
the Einstein relation between the TP mobility and the diffusivity
generally holds, despite the fact that in some cases diffusion
is anomalous. 
For models treated within the framework of the decoupling approximation,
our analytical results are confirmed by Monte Carlo simulations.
Perturbance of the lattice gas particles distribution 
due to the presence of a biased TP
and the form of the particle density profiles are also discussed.

\end{abstract}

\end{center}

\end{titlepage}

\section{Introduction.}

Transport in media with quenched or time-dependent disorder 
has received much interest 
due to the considerable fundamental and technological 
importance of the problem \cite{1,2,3,4,44,444,4444,5}. Transport processes 
underlay a variety of chromatographic separation techniques \cite{4,44,444}, 
extraction of oil from porous rocks \cite{4,44,444}, excitations quenching in 
amorphous solids and doped crystals \cite{444,4444,6,7,8,9}, conductivity of 
complex media and permeability of disordered membranes \cite{5,10}.
Molecular diffusion measurements in disordered media often serve 
as a tool in characterizing the internal geometry over 
a range of molecular and macroscopic length scales \cite{4,44,444,4444}.
In addition, transport processes control the kinetics of
chemical reactions in disordered media \cite{6,7,8,9}. 

Following motion of individual particles in disordered
media one most often 
encounters behavior characterized by a mean-square displacement 
$<{\bf R}^2(t)>$
growing as
\begin{equation}
\label{dif}
<{\bf R}^2(t)> \; \sim \; D \; t,
\end{equation}
i.e. linearly in time with 
$D$ being the "diffusion" coefficient, 
which 
depends on the peculiar features 
of the disordered medium. 
Such type of dynamics is ubiquitous due to its inherent relationship
to the central-limit theorem, which makes it a paradigm of stochastic
processes. 

One the other hand, there was a growing evidence for other laws of
random transport, both slower and enhanced relative to the time dependence
in Eq.(\ref{dif}). Generally, one finds 
\begin{equation}
\label{andif}
<{\bf R}^2(t)> \sim t^{\alpha}, \;\;\; \text{with} \;\;\; \alpha \neq 1.
\end{equation} 
Case with $\alpha < 1$, known as the dispersive transport or subdiffusive
regime, has been in the focus of many theoretical  
and experimental studies (see, e.g., Refs.\cite{1,2,3,4,9}) and 
has been attributed, for instance, to
random walks on fractal structures \cite{1,2,3,4,9} or to 
diffusion in presence of 
temporal traps \cite{1,9,11}. A few stray examples of such a transport are
time of flight measurements in amorphous semiconductors \cite{11,12},
defect diffusion in relaxation studies \cite{11}, excitation migration
on the disordered array of donor centers (see, e.g., Ref.\cite{122}), 
tracer 
diffusion in a linear array of convective rolls \cite{13}
or in low-dimensional lattice gases \cite{14,lev,per,15,16,17}, or
the so-called Sinai diffusion with extremely confined particle trajectories
such that $<{\bf R}^2(t)> \sim \ln^4(t)$ \cite{18,19,20}.
Enhanced diffusion regime, $\alpha > 1$, is observed in turbulent diffusion 
($\alpha = 3$), as well as in 
some other instances (see, e.g., Ref.\cite{21}) for some review). 

Suppose now that in the disordered medium in question, in which, in absence
of any fields acting on the species whose motion we follow (say, tracer particles),
we find that their 
mean displacement is zero, while the
mean-square displacement 
$<{\bf R}^2(t)>$ obeys Eq.(\ref{dif})  or the anomalous diffusion law in Eq.(\ref{andif}).
Suppose next that we manage to 
bias somehow the particle motion, 
say, by charging the tracer particles only and 
putting the whole system in a constant 
electric field $E$. This situation 
is, of course, quite interesting in its own right and is
appropriate
to charge transfer or "dynamic directed 
percolation" in time-dependent 
inhomogeneous medium \cite{22,23,24,25}, such as, e.g., certain biomembranes \cite{40},  
solid protonic conductors \cite{50}, 
oil-continuous microemulsions \cite{60,70,80,90} or polymer
electrolytes \cite{100,110,1200}. 
In such a situation, the tracer particles will move 
preferentially in the direction of the field
and will attain a non-zero mean displacement 
and a non-zero mean velocity $V(t)$. 
One may define then the time-dependent 
tracer particle mobility as
\begin{equation}
\label{mob}
\mu(t) = \lim_{E \to 0} \frac{V(t)}{E},
\end{equation}
while the "diffusivity" 
$D(t)$ 
can be extracted from the behavior in $absence$ of the 
field (Eqs.(\ref{dif}) or (\ref{andif})) via:
\begin{equation}
\label{Di}
D(t) = \frac{< {\bf R}^2(t) >}{2 d t},
\end{equation}
where $d$ is the spatial dimension of the system.

Now, the question is whether the mobility $\mu(t)$, 
calculated from the tracer particle 
mean displacement in the presence of an external
electric field, and the diffusivity  $D(t)$, Eq.(\ref{Di}), deduced from the 
tracer particle 
mean-square displacement in the absence of
the field, obey the generalized Einstein relation of the form
\begin{equation}
\label{einstein}
\mu(t) = \beta D(t),
\end{equation} 
where $\beta$ denotes the reciprocal temperature?

The answer is trivially positive, of course, 
for isolated non-interacting particles performing conventional Brownian motion.
For particle diffusion in disordered medium the validity of the relation in
Eq.(\ref{einstein}) is much less evident. 
On the one hand, it has been found that 
Eq.(\ref{einstein}) holds 
for the tracer particle 
motion in a one-dimensional (1D) gas of hard spheres, whose whole
dynamics consists of pairs of neighboring particles interchanging velocities at each
collision and the initial velocity distribution being an equilibrium one \cite{lp}.
It holds also for the tracer particle
diffusion in a 1D hard-core gas 
with diffusional dynamics
when the 1D
lattice is a ring of a finite length 
\cite{lebowitz2},
as well as
 for infinite 1D lattices with non-conserved 
\cite{benichou} and conserved particles number
\cite{bur2,olla,bur3}. 
Remarkably, in the latter 
case Eq.(\ref{einstein}) holds 
for $t$ sufficiently large despite the fact that both 
$\mu(t)$ and $D(t)$ 
do not attain constant values
as $t \to \infty$ but rather are decreasing functions of time
\cite{bur2,olla,bur3}. 
In addition, 
the validity of the Einstein relation 
has been corroborated 
for the charge carriers in semiconductors \cite{gu} and 
for polymeric systems in the subdiffusive regime \cite{am,ba}. 

On the other hand, it is well known that the Einstein 
relation is violated in some physical situations; 
for instance, it is
not fulfilled 
for Sinai diffusion \cite{18,19,20} 
describing Brownian motion of a test particle in presence of a quenched
random force, and for the Scher-Lax-Montroll model of
anomalous random walk \cite{barkai} (see also Refs.\cite{cugl1} 
and \cite{cugl2} for some other examples). 
Apparently, the Einstein relation is also 
violated for random walk
in disordered lattices at their percolation threshold
- the celebrated model of the "ant in the labyrinth"
of de Gennes \cite{pgg}. Here, in absence of external 
force acting on the random walker, its mean-square 
displacement $<X^2(t)>$ obeys Eq.(\ref{andif}) with $\alpha < 1$
\cite{pgg,gef}. In presence of external field, which produces a 
bias in the random walk making the walker more 
likely to step along the field than
against it, the behavior appears to be rather complex. As a matter of fact, the bias
has a dual effect - it induces drift in the direction of the 
field, but also creates temporal traps, such as in dead-end branches, 
to escape from which the walker
has to move against the field. One then finds that the drift velocity is not a monotonic function
of the bias 
\cite{bot,bar} and moreover, it has been argued that the drift velocity
vanishes
once the bias exceeds a certain threshold value \cite{bar} (see also \cite{dhar}). 
These theoretical arguments have been supported by 
exact calculation of the mean drift velocity
on a random comb and randomly diluted Bethe lattice \cite{dhar2} (see, however, Ref.\cite{gef2} for some objections).
Therefore, a small value of external bias gives rise to a 
mean displacement in the direction of the field that increases linearly with time, and the mean velocity tends to a constant.
This asymptotic value of the mean velocity is proportional to the 
field for small fields \cite{dhar}, which
implies that the mobility tends to a constant as $t \to \infty$. 
Contrary to that, 
since in the absence of the field the walker motion is subdiffusive \cite{pgg,gef}, 
one has that here the diffusivity $D(t)$ should vanish as $t \to \infty$.
Therefore, if the results of Refs.\cite{bot,bar,dhar,dhar2} are correct, 
they would imply
that the
Einstein relation is not valid
 for diffusion in random lattices at percolation threshold.

Hence,    
in principle, 
 it is not $\it a \; priori$ clear at all
whether  Eq.(\ref{einstein}) should be valid in any case. 

In this presentation we overview some recent results on 
the generalized Einstein relation in Eq.(\ref{einstein})
for tracer particle (TP) 
diffusion in dynamical disordered environments, as exemplified here by 
non-interacting lattice gases of hard-core particles.  
The paper is structured as follows: 
In Section 2 we introduce the model, basic notations, write down 
dynamic equations describing the time evolution of the system
and discuss 
the way of their solution. In Section 3 we first 
recall the 
properties of an unbiased TP diffusion in 
a 1D hard-core lattice gas. Next, we turn to the 
case of a biased TP
in an inert lattice gas and present explicit results describing the TP mean 
displacement and the mobility. 
We demonstrate then that the Einstein relation in Eq.(\ref{einstein})
holds exactly despite the fact that in this case diffusion is anomalous. 
Further on, in Section 4 we consider the case 
when the initial particle
distribution on a 1D 
lattice is inhomogeneous and 
characterized by an $S$-shape, "shock"-like 
density 
profile. We find that when the TP is attracted towards
the particle phase with higher density, at a certain 
critical value of the constant 
attraction force
the TP will
not move, on average. Assuming the validity of the Einstein relation,
we find the second moment of the TP displacement. This heuristic 
analytical result
is confirmed numerically
which signifies that the Einstein relation holds apparently for such 
an inhomogeneous situation.
Next, in Sections 5 and 6 we     
analyse the forms of the biased 
TP terminal velocity 
in one-, two- and three-dimensional  
hard-core lattice gases undergoing 
continuous particles exchanges with
a reservoir. 
We determine the TP mobility and assuming the validity of the Einstein relation,
obtain the TP diffusion coefficient $D$ 
in the unbiased case. Results for 1D systems 
are confirmed by numerical Monte Carlo simulations. 
We also show that in two- and three-dimensional systems 
in the conserved particle density limit (when particle exchanges 
with the reservoir are forbidden)
our predictions for $D$ coincide with classical results of 
Nakazato and Kitahara \cite{nakazato}, 
which are known to be in a very good agreement with the numerical data \cite{kehr}.
In Section 7 we present an exact solution for the biased 
TP
mean displacement in a two-dimensional (2D) lattice gas 
containing a single unoccupied site.
On comparing our results with earlier results by 
Brummelhuis and Hilhorst \cite{hilhorst}
obtained for the unbiased 
TP dynamics, 
we infer that also in this case, in which the TP diffusion is anomalously, 
logarithmically confined,
the Einstein relation does hold exactly. Finally, we conclude in Section 
8 with a brief summary of results and discussion.   

\section{The model and basic equations.}

Consider a
$d$-dimensional 
regular hypercubic
lattice of spacing $\sigma$ each site of which
is brought 
in
contact with a reservoir containing identic,
electrically neutral particles  - a vapor phase (Fig.1), 
maintained at a constant pressure. 
We suppose next that the reservoir particles 
may be created (adsorbed) 
on any vacant site at a fixed rate $f/\tau^*$, which rate 
depends on the reservoir
pressure and the energy gain due to the "creation"  event. Further on,
the particles may move randomly along the lattice by  
hopping at a rate $1/2 d \tau^*$ to any of $2 d$
neighboring lattice sites,
which process is 
constrained by hard-core 
exclusion preventing multiple occupancy of any of the sites.
Lastly, the particles may 
spontaneously disappear  (desorb) 
from the lattice 
at rate $g/\tau^*$, which is dependent on the barrier against desorption. 
Both $f$ and $g$ are site and environment independent.

\begin{figure}[h]
\begin{center}
\includegraphics*[scale=0.4]{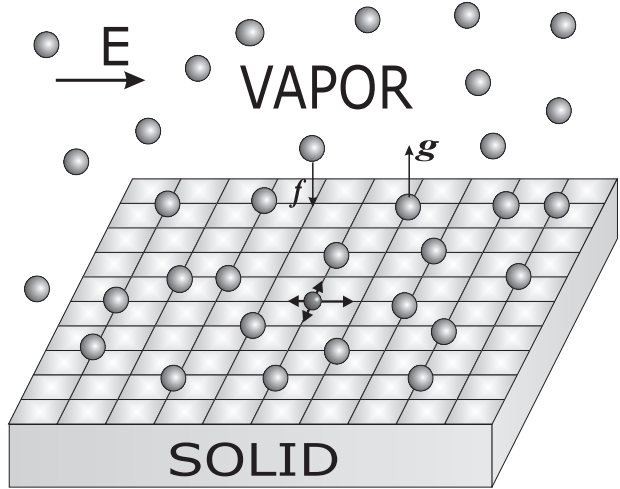}
\end{center}
\caption{\label{reseau}\small \sl Two-dimensional realization of the model:
Adsorbed monolayer in contact with a vapor phase. 
Gray spheres denote the
monolayer (vapor) particles; the smaller black sphere 
stands for the driven tracer
particle.}
\end{figure}

 To describe the occupancy of lattice sites, we introduce 
a time-dependent 
variable $\eta({\bf R})$, which may assume two values:
\begin{equation}
\eta({\bf R}) = \left\{\begin{array}{ll}
1,     \mbox{ if the site ${\bf R}$ is occupied,} \nonumber\\
0,     \mbox{ if the site ${\bf R}$  is empty.}
\end{array}
\right.
\end{equation} 
Note that the local variable $\eta({\bf R})$ can 
change its value due to creation of particles (adsorbtion),
desorption and random hopping events. 
Note also that random hopping events do conserve the total number of adsorbed
particles and hence, the average particle 
density $\rho_s(t)$. On the other hands, 
creation/desorption processes do change $\eta({\bf R})$ 
locally such that   
the total number of 
particles on the lattice 
is not explicitly conserved when such two processes are allowed. 
However, the mean particle density, $\rho_s(t) = <\eta({\bf R})>$, approaches as
$t\to\infty$ a constant value
\begin{equation}
\rho_s=\frac{f}{f+g}
\label{rhostat}
\end{equation} 
This  relation is well known and represents the customary Langmuir adsorption
isotherm \cite{surf}. We finally remark that in the analysis of the stationary-state behavior,
 we can always turn to the conserved particles number limit by 
setting $f$ and $g$ equal to zero
 and keeping their ratio
fixed, i.e. supposing that $f/g = \rho_s/(1 - \rho_s)$. 
This limit will correspond to the model of biased tracer diffusion in
a  hard-core lattice gas with fixed particles density $\rho_s$, 
and will allow us to check our analytical 
predictions against some already known results
\cite{kehr,nakazato,hilhorst,tahir,beijeren}. 
  
Now, we introduce a bias into the system. Contrary to the well-studied situation
in which all particles are subject to a constant force  
(see, e.g., Refs.\cite{katz,satya1,satya2,aro} and references therein), we will focus
here on essentially less studied case when only one of particles is subject to external bias
while the rest have symmetric hopping probabilities. This particle may be thought of as a
probe designated to measure the frictional properties of symmetric non-inteacting lattice gas.
To this end, at $t = 0$ we introduce at the lattice origin 
an extra hard-core 
particle, whose motion
we would like to follow and which will be called in what follows as the tracer
particle - the TP;  
position of this particle at 
time $t$ is denoted as ${\bf R}_{tr}$. 
We stipulate that the TP is different from other particles in two aspects:
a) it can not disappear from the lattice and b)  
it is subject to
some external driving force, which favors its jumps into a
preferential direction. 
Physically, (a) be realized, 
for instance,
if this only particle is charged and the system is subject to a
uniform electric field {\bf E}.  We suppose here, 
for simplicity of exposition, that the external force ${\bf E}$
 is oriented according to the unit vector 
${\bf e_1}$.

The dynamics of the biased TP is defined in the usual fashion \cite{17}:
We suppose
that the TP, which occupies the site ${\bf R}_{tr}$ at time $t$,
 waits an exponentially distributed time with mean
$\tau$, 
and then attempts to hop onto one of $2 d$ neighboring sites, ${\bf
R}_{tr} + {\bf e}_{\boldsymbol \nu}$, 
where ${\bf e}_{\boldsymbol \nu}$
are $2 d$  unit 
vectors of the hypercubic lattice. 
In what follows we adopt the notation $\nu = \{\pm1,\pm2\, \ldots ,\pm d\}$, where 
${\bf  e_1}$  will denote 
the  direction of the external force ${\bf E}$.  Next, the jump direction is chosen 
according to the probablity $p_\nu$, which obeys:
\begin{equation}
p_\nu=\frac{\exp\Big[\frac{\beta}{2}({\bf E \cdot e}_{\boldsymbol
\nu})\Big]}{\sum_{\mu}\exp\Big[\frac{\beta}{2}({\bf E \cdot e}_{\boldsymbol \mu})\Big]},
\label{defp}
\end{equation}    
where $\beta$ is the reciprocal temperature, $({\bf E \cdot e})$ stands for the scalar product, 
the charge of the TP is set equal to unity and  
the sum with the subscript $\mu$ denotes summation over all possible
orientations
of the vector ${\boldsymbol e_\mu}$; that is, 
$\mu = \{\pm1,\pm2\, \ldots , \pm d\}$. 

After the jump direction is chosen, the TP attempts to hop 
onto the target site. The hop is instantaneously fulfilled 
if 
the target site
is vacant at this moment of time; otherwise, i.e., if the target site
is occupied by any adsorbed particle, the jump is rejected and the
TP remains at its position.

To close this subsection, several comments on the relevance of the model
to some physical situations are in order. As we have already remarked, in the
conserved particles number limit the model under study reduces to the model of biased
tracer diffusion in hard-core lattice gases, which is related to a wide variety of
physical situations, including, for example, diffusion of interstitials in crystals. 
A detailed review can be found in Ref.\cite{kehr}.  In two-dimensions, in the
non-conserved particles number case it describes tracer diffusion in adsorbed
monolayers in contact with the vapor phase \cite{benichouPRB} and allows to determine,
for instance, the analog of the Stokes 
formula for such monolayers and
consequently, to elucidate their intrinsic frictional properties.
On the other hand, the 2D case can be thought of as some simplified
picture of the stagnant layers emerging in liquids that are in contact with a solid
body. It is well-known (see, e.g. Ref.\cite{lyklema}) that liquids in the close
vicinity of a solid interface - at distances within a few molecular diameters - do
possess completely different physical properties compared to those of the bulk phase.
In this stagnant region, in which an intrinsically disordered liquid phase is spanned
by and interacts with the ordering potential of the solid, the liquid's viscosity is
drastically enhanced and transport processes (related, say, to biased diffusion of
charged carriers in solutions) are substantially hindered. Thus our model can be
viewed as a two-level approximate model of this challenging physical system, in which
the reservoir mimics the bulk fluid phase with very rapid transport, while the
adsorbed monolayer (particles on a 2D lattice) represents the stagnant layer emerging
on the solid-liquid interface. Lastly, the model under study represents a novel
example \cite{25} of the so-called "dynamic percolation" models (see,
e.g., Refs.\cite{22,23,24} and references therein), invoked to describe transport
processes in many situations with dynamical disorder. In this context, the lattice gas
particles can be thought of as representing some fluctuating environment, which
hinders the motion of an impure molecule - the TP, which might be, for example, a
charge carrier. A salient feature of this model, which makes it different from the
previously proposed models of dynamic percolative environments, is that we include the
hard-core interactions between the "environment" particles and the TP, such that the
latter may itself influence the dynamics of the environment.  As we proceed to show, this circumstance
turns out to be very important and will result in formation of inhomogeneous 
distribution of the environment particles.

\subsection{Evolution equations.}

Now, we derive the evolution equations in a general, $d$-dimensional case.  
 We begin by introducing some auxiliary definitions.
Let $\eta\equiv
\{\eta({\bf R})\}$ denote the entire set of the occupation variables, which defines the instantaneous
configuration of the lattice-gas  particles  at time moment $t$. Next, let 
$P({\bf R_{tr}},\eta;t)$ stand for the joint probability of finding  
at time $t$ the TP at the site ${\bf R_{tr}}$
and all adsorbed particles in the configuration $\eta$.
Then, denoting as $\eta^{{\bf r},\nu}$ a
configuration obtained from $\eta$ by the Kawasaki-type exchange of the occupation
variables of two neighboring 
sites ${\bf r}$ and ${\bf r+e}_{\boldsymbol \nu}$,
and as 
${\hat \eta}^{\bf r}$ - a configuration obtained from 
the original  $\eta$ by
the replacement $\eta({\bf r}) \to 1-\eta({\bf r})$, which corresponds to the
Glauber-type flip of the occupation variable due to 
the creation/desorption 
events, we have that 
the time evolution of the configuration
probability $P({\bf R_{tr}},\eta;t)$ obeys the following master equation:
\begin{eqnarray}
&&\partial_tP({\bf R_{tr}},\eta;t)=
\frac{1}{2d\tau^*}\sum_{\mu=1}^d\;\sum_{{\bf r}\neq{\bf R_{tr}}-{\bf e}_{\boldsymbol \mu},{\bf R_{tr}}}  \; 
\Big\{ P({\bf R_{tr}},\eta^{{\bf r},\mu};t)-P({\bf R_{tr}},\eta;t)\Big\}\nonumber\\
&+&\frac{1}{\tau}\sum_{\mu}p_\mu\Big\{\left(1-\eta({\bf R_{tr}})\right)P({\bf R_{tr}}-{\bf e}_{\boldsymbol \mu},\eta;t)
-\left(1-\eta({\bf R_{tr}}+{\bf e}_{\boldsymbol \mu})\right)P({\bf R_{tr}},\eta;t)\Big\}\nonumber\\
&+&\frac{g}{\tau^*}\sum_{{\bf r}\neq {\bf R_{tr}}} \;\Big\{\left(1-\eta({\bf r})\right)P({\bf R_{tr}},
\hat{\eta}^{{\bf r}};t)-\eta({\bf r})P({\bf R_{tr}},\eta;t)\Big\}\nonumber\\
&+&\frac{f}{\tau^*}\sum_{{\bf r}\neq{\bf R_{tr}}} \;\Big\{\eta({\bf r})P({\bf R_{tr}},\hat{\eta}^{{\bf r}};t)
-\left(1-\eta({\bf r})\right)P({\bf R_{tr}},\eta;t)\Big\}.
\label{eqmaitresse}
\end{eqnarray} 

The mean 
velocity $V(t)$ 
of the TP can be obtained by multiplying both sides of Eq.(\ref{eqmaitresse}) by
$({\bf R_{tr} \cdot e_1})$ and summing over all possible configurations $({\bf R_{tr}},\eta)$.
This results in the following exact equation determining the TP velocity:
\begin{equation}
V(t) \equiv\frac{d}{dt} \; \sum_{{\bf R_{tr}},\eta} ({\bf R_{tr} \cdot e_1})P({\bf R_{tr}},\eta;t) =
\frac{\sigma}{\tau}\Big\{p_1  \Big(1-k({\bf e_1};t)\Big)-p_{-1} \Big(1-k({\bf e_{-1}};t)\Big)\Big\},
\label{vitesse}
\end{equation}
where 
\begin{equation}
k({\boldsymbol \lambda};t)\equiv\sum_{{\bf R_{tr}},\eta}\eta({\bf R_{tr}}+{\boldsymbol \lambda})P({\bf R_{tr}},\eta;t)
\label{defk}
\end{equation}
is the probability of having at time t a lattice gas 
particle 
at position ${\boldsymbol \lambda}$, 
defined in the frame of reference moving with the TP. 
In other words,  $k({\boldsymbol \lambda};t)$
 can be thought of as being 
the particle 
density
profile as seen from the moving TP. 

Equation (\ref{vitesse}) signifies that 
the velocity of the TP is 
dependent on the lattice gas particles 
density in the 
immediate vicinity of the tracer. 
If the lattice gas is perfectly stirred, or, in other words, if 
$k({\boldsymbol \lambda};t) = \rho_s$ everywhere, (which implies immediate 
decoupling of ${\bf R_{tr}}$ and $\eta$),  one would obtain
from Eq.(\ref{vitesse}) a trivial mean-field result
\begin{equation}
V^{(0)}=(p_1-p_{-1})(1-\rho_s)\frac{\sigma}{\tau},
\label{vmf}
\end{equation}
which states that the only effect of the medium on the TP dynamics is that its 
jump time $\tau$ is merely renormalized by a
factor $(1 - \rho_s)^{-1}$; 
$(1 - \rho_s)/\tau $ defines simply 
the mean frequency of successful jump events. 

However, the situation appears to be more complicated and, as we proceed to show, 
$k({\boldsymbol \lambda};t)$ is 
different from the equilibrium value $\rho_s$ everywhere, except for 
$|\boldsymbol \lambda|\to\infty$. This means that the TP strongly 
perturbs the particles distribution on the lattice - it is no longer
uniform and some non-trivial density profiles emerge. 

Now, in order to calculate the instantaneous mean
velocity of the TP we have to determine the mean particles density 
at the neighboring to the TP sites
${\bf R_{tr}}+{\bf e_{\pm1}}$, which requires, in turn, computation of 
the density profile $k({\boldsymbol \lambda};t)$ for arbitrary $\boldsymbol \lambda$.
The latter can be found
 from the master equation     
(\ref{eqmaitresse}) by multiplying both sides 
by $\eta({\bf R_{tr}})$ and performing the summation over all
configurations $({\bf R_{tr}},\eta)$. In doing so, we find 
that these equations are not closed with respect to 
$k({\boldsymbol \lambda};t)$, 
but are coupled to the third-order
correlations,
\begin{equation}
T({\boldsymbol \lambda},{\bf e}_{\boldsymbol \nu};t) =\sum_{{\bf R_{tr}},\eta}\eta({\bf R_{tr}}+{\boldsymbol \lambda})\eta({\bf R_{tr}}+{\bf e}_{\boldsymbol \mu})P({\bf R_{tr}},\eta;t)
\end{equation}
In turn, if we proceed further to
the third-order
correlations, we find that these are 
coupled respectively to the fourth-order correlations. 
Consequently, in order
to compute $V(t)$, one faces
the problem of solving an infinite hierarchy of coupled equations.  

We resort then to the simplest non-trivial 
closure of the hierarchy    
in terms of $k({\boldsymbol \lambda};t)$, which has been first proposed in
Ref.\cite{bur2}, and  represent $T({\boldsymbol
\lambda},{\bf e}_{\boldsymbol \nu};t)$ 
as
\begin{eqnarray}
&&\sum_{{\bf R_{tr}},\eta}\eta({\bf R_{tr}}+{\boldsymbol \lambda})\eta({\bf R_{tr}}+{\bf e}_{\boldsymbol \mu})P({\bf R_{tr}},\eta;t)\nonumber\\
&\approx&\left(\sum_{{\bf R_{tr}},\eta}\eta({\bf R_{tr}}+{\boldsymbol \lambda})P({\bf R_{tr}},\eta;t)\right)\left(\sum_{{\bf R_{tr}},\eta}
\eta({\bf R_{tr}}+{\bf e}_{\boldsymbol \mu})P({\bf R_{tr}},\eta;t)\right)\nonumber\\
&=&k({\boldsymbol \lambda};t)k({\bf e}_{\boldsymbol \mu};t), 
\label{decouplage}
\end{eqnarray}
Some arguments justifying such an approximation can be found in
Ref.\cite{benichou}. 

Using  the approximation in Eq.(\ref{decouplage}), we obtain
\begin{equation}
2d\tau^*\partial_tk({\boldsymbol \lambda};t)=\tilde{L}k({\boldsymbol \lambda};t)+2df, \;\;\; \tilde{L}\equiv\sum_\mu A_\mu\nabla_\mu-4(f+g),
\label{systemek1}
\end{equation}
which holds for all ${\boldsymbol \lambda}$, except
 for  ${\boldsymbol \lambda}=\{{\bf 0},\pm{\bf e_1},{\bf e}_{2}\ldots,{\bf e_d}\}$. One the
other hand, for these special sites ${\boldsymbol \lambda} = {\bf e_{\nu}}$ 
with $\nu=\{\pm1,2,\ldots, d\}$ we find
\begin{equation}
2d\tau^*\partial_tk({\bf e}_{\boldsymbol \nu};t)=(\tilde{L}+A_\nu)k({\bf e}_{\boldsymbol \nu};t)+4f,  
\label{systemek2}
\end{equation}
where 
the coefficients $A_{\mu}$ are defined by
\begin{equation}
A_\mu(t)\equiv1+\frac{2d\tau^*}{\tau}p_\mu(1-k({\bf e}_{\boldsymbol \mu};t)).
\label{defA}
\end{equation}
Note that Eq.(\ref{systemek2}) represents, from the mathematical point of view, 
the boundary conditions for the general evolution equation  (\ref{systemek1}), 
imposed on the sites in the
immediate vicinity of the TP. Equations (\ref{systemek1}) and (\ref{systemek2}) 
together with Eq.(\ref{vitesse}) thus constitute
a closed system of equations which suffice computation of all properties of 
interest, i.e. the time-dependent tracer particle velocity
and the particle density profiles as seen from the tracer particle. 
The general approach to solution of coupled non-linear 
Eqs.(\ref{vitesse}),(\ref{systemek1}) and (\ref{systemek2})
has been discussed in detail in  Ref.\cite{benichouPRB}. 
Here we merely note that 
despite the fact that using the decoupling scheme in
Eq.(\ref{decouplage}) we effectively close the system of 
equations on the level of the pair
correlations, solution of   Eqs.(\ref{systemek1}) and
(\ref{systemek2}) still 
poses serious technical 
difficulties: Namely, 
these equations are non-linear with respect to the TP velocity, 
which enters the gradient term on
the rhs of the evolution equations for the pair correlation, and 
does depend itself on the values of the
monolayer particles densities in the immediate vicinity of the TP. 
Solution of this
system of non-linear equations for several particular case
will be presented in the  next sections.

\section{Biased tracer diffusion in a 1D lattice gas.}

Consider first biased tracer diffusion in the simplest case of a 
1D hard-core lattice gas with conserved
particles number; that is, a lattice gas in which all particles were 
initially introduced
onto the lattice and particle exchanges with the vapor phase are forbidden ($f = g = 0)$. 
Hence, the particle density
$\rho_s$ on the lattice is kept constant all the time. For simplicity, we 
also set here the spacing $\sigma = 1$ and $\tau = \tau^* = 1$. These parameters can be trivially
restored in final results. 

In absence of external field acting on the TP, when its transition probabilities are symmetric,  
tracer dynamics has been studied first 
in Refs.\cite{14,lev,per,15,16} and the 
TP mean-square displacement $< X^2(t)>$ has been determined exactly:
\begin{equation}
\label{harr}
<X^2(t)> = \frac{(1 - \rho_s)}{\rho_s} \sqrt{\frac{2 t}{\pi}},
\end{equation}
which thus follows an anomalous diffusion law in Eq.(\ref{andif}) with $\alpha = 1/2$. Consequently,
the "diffusivity" D(t) is given here by
\begin{equation}
\label{harris}
D(t) = \frac{(1 - \rho_s)}{\rho_s} \frac{1}{\sqrt{2 \pi t}}  \to 0 \;\;\; \text{as} \;\;\; t \to \infty.
\end{equation}

Now, if we charge the TP and switch on an external electric field, how will
the TP velocity and the mobility behave? An early incomplete answer
was presented in Ref.\cite{burbur} which focused on the  
extreme case of an infinitely strong electric field
$E = + \infty$ such that the TP performs 
a totally directed random walk
constrained by hard-core particles. 
By noticing that such a model
is tantamount to the 1D
Stefan freezing problem or 
the problem of directional solidification \cite{crank}, (the 
role of the latent heat being played by the
lattice gas particles), it was found
that in this extreme case the TP mean velocity 
follows
\begin{equation}
\label{speed}
V(t) = \frac{\displaystyle \gamma_\infty}{\sqrt{t}}, 
\end{equation}
where $\gamma_\infty$ is a function of $\rho_s$
defined implicitly by
\begin{equation}
\sqrt{\frac{\pi}{2}} \gamma_\infty \exp\Big(\frac{\displaystyle \gamma^2_\infty}{2}\Big)
{\rm erfc}(\frac{\gamma_\infty}{\sqrt{2}}) = 1 - \rho_s, 
\end{equation}
${\rm erfc(x)}$ being the complementary error function.
Hence, the TP mean velocity in the biased case 
decays
at exactly the same rate
as the TP "diffusivity" in the unbiased case! In view of such a coincidence, 
one is, of course, prompted to consider the general case $E < \infty$
and to check the prefactors in Eq.(\ref{speed}) having in mind to verify the validity of 
Eq.(\ref{einstein}) for anomalous 
tracer diffusion in such random dynamical environments  \cite{qq}. 
 
The general case of arbitrary $0 < E \leq \infty$ 
has been studied in Ref.\cite{bur2} within the framework 
of the approach outlined in Section 2 and subsequently, 
in terms of a rigorous probabilistic approach in Ref.\cite{olla}.
Results of these two approaches coincide, which justifies 
the decoupling approximation in Eq.(\ref{decouplage}).   
It has been found that in this general case the TP mean velocity  
$ V(t) $ obeys at sufficiently large times 
the form of Eq.(\ref{speed}), i.e. $V(t) = \gamma_E/\sqrt{t}$, in which law
the prefactor $\gamma_E$
is determined by the following equation \cite{bur2,olla}:
\begin{eqnarray}
\label{k}
&&\left(I_+(\gamma_E) - 1 + \frac{\rho_s}{1 - \exp(- \beta E)}\right) \times \nonumber\\
&\times& \left(I_-(\gamma_E) + 1 + \frac{\rho_s}{\exp(\beta E) - 1}\right) =
\frac{\exp(\beta E) \rho_s^2}{\displaystyle\Big(\exp(\beta E) - 1 \Big)^2},
\end{eqnarray}
where
\begin{equation}
I_{\pm}(\gamma_E) = \sqrt{\frac{\pi}{2}} \gamma_E \exp\Big(\frac{\displaystyle \gamma^2_E}{2}\Big)
 \left[1 \mp erf\Big(\frac{\gamma_E}{\sqrt{2}}\Big)\right]
\end{equation}
In the limit $\beta E \ll 1$ the solution of the transcedental Eq.(\ref{k}) has the form
\begin{equation}
\gamma_E \approx \frac{(1 - \rho_s)}{\rho_s} \sqrt{\frac{2}{\pi}} \frac{\beta E}{2},
\end{equation}  
which implies that the time-dependent mobility in this case is given by
\begin{equation}
\mu(t) = \frac{(1 - \rho_s)}{\rho_s} \frac{\beta}{\sqrt{2 \pi t}}.
\end{equation} 
Finally, by comparing the prefactors in the latter equation and Eq.(\ref{harris}), we find
that 
the generalized Einstein relation in Eq.(\ref{einstein}) 
holds exactly for such an anomalous diffusion!

To close this section, we make several comments on the particle density 
distribution as seen from the tracer particle.
Since the TP has a non-zero mean displacement, 
while the lattice gas particles have 
symmetric hopping
probabilities and hence, a zero mean displacement, 
they tend to accumulate in front of
the TP.   Consequently, a condensed, traffic jam-like region is formed 
in front of the TP, and a depleted by the lattice-gas 
particles region past the TP emerges. 
As a matter of fact, 
such a inhomogeneous density distribution is not stationary: the size of the
traffic jam-like region, in which the particle density profile $k(\lambda,t)$ is nearly 
constant and larger than $\rho_s$, $k(\lambda,t) = \rho_s/(1 + I_+(\gamma_E))$, 
grows in proportion to the TP mean
displacement. The size of the depleted region 
with a constant density $k(\lambda,t) = \rho_s/(1 + I_-(\gamma_E)) < \rho_s$ also grows in proportion
to $<X(t)> = V(t) t$.

\section{"Shock" propagation in a non-equilibrium 1D lattice gas.}

In this section we address the question of the validity of the generalized Einstein relation
in a somewhat exotic situation in which the initial 
particle distributions from both sides of the 
TP has an $S$-shape, shock-like profile 
characterized by two different mean densities \cite{bur3}. That is, as depicted in Fig.2,
we suppose that 
the TP is initially at the origin, and the particles mean densities 
from the left and from the right of the origin, which we denote as
$\rho_-$ and $\rho_+$, respectively, are not
equal to each other. The TP thus defines position of the "shock front". 
Without of lack of generality, we suppose that  $\rho_- \geq \rho_+ \geq 0$, 
and will call in what follows
the particle phase which initially occupied the left half-line as the 
high-density phase (HDP), while the phase initially
occupying the right half-line will be referred to as the low-density phase (LDP).

\begin{figure}[h]
\begin{center}
\includegraphics*[scale=0.8]{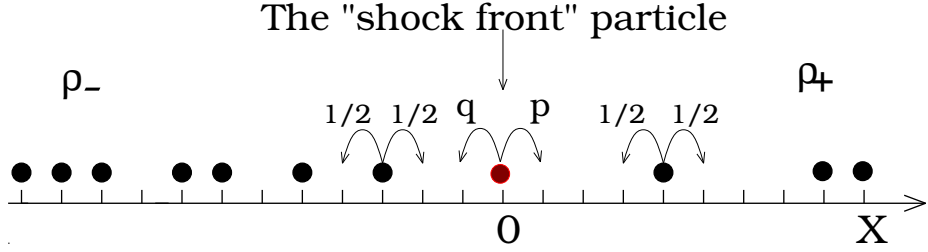}
\end{center}
\caption{\label{re}\small \sl 
Initial "shock"configuration of the lattice gas particles on a 
1D lattice. $\rho_-$ is the initial mean density of particles at 
the half-line $-\infty < X < 0$ and $\rho_+$ is the mean density of particles at the half-line $0 < X < \infty$.
All lattice gas particles, excluding the TP (the "shock" front particle), have symmetric transition probabilities. The TP has
asymmetric transition probabilities and experiences the bias in the negative direction.}
\end{figure}  

Further on, after deposition on the lattice, particles start to perform
$symmetric$, (i.e. with equal probabilities for going to the 
left or to the right),
hopping motion
between the 
nearest lattice sites under the
constraint that neither two
particles can simultaneously
occupy the
same lattice site and can not
pass through each other. 
Next, we stipulate that the TP is subject to
a constant force $F$ which favors its jumps 
in a preferential direction. 
Thus for the TP the probabilities of
going to the right ($p$) and to the left ($q$) will be different 
from each other.  We will suppose that $F$ is directed towards
the HDP. We adopt the convention that in this case
$F$ is negative, $F \leq 0$; the TP hopping probabilities
$p$ and $q$ 
are related to the force 
and the reciprocal temperature $\beta$
 through $p/q = exp(\beta F)$ and
$p + q = 1$.
 From the physical point
of view, such a $constant$ force can be understood
as an effective boundary tension  derived from the
solid-on-solid-model Hamiltonian of the
 phase-separating boundary 
 \cite{joela,joelb,joelc} and 
mimic, in a mean-field fashion, the presence of
attractive interactions
between the lattice-gas particles  
which are not explicitly included in the model. Relation between $p$, $q$
and the amplitude of attractive interparticle 
interactions 
has been discussed in Refs.\cite{ya} and \cite{ya1}.

The TP dynamics in such a system has been analysed in Refs.\cite{bur3}
and \cite{olla} and it has been found that the TP mean velocity obeys
$< V(t) > = \gamma_F/\sqrt{t}$ at sufficiently large times, where 
the prefactor $\gamma_F$ is determined implicitly as the solution of the following equation:
\begin{equation}
\label{j}
\frac{q \; \rho_{-}}{1 + I_{-}(\gamma_F)} \; - \; \frac{p \; \rho_{+}}{1 - I_{+}(\gamma_F)} \; = \; q \; - \; p,
\end{equation}
which generalizes the result of the previous section 
for the case of $\rho_{-} \neq \rho_{+}$.

Equation (\ref{j}) predicts that three
different regimes can take place depending on the relation
between $p/q$ and $\rho_{\pm}$:

(1)  When $p (1 - \rho_{+}) > 
q ( 1 - \rho_{-})$ the parameter $\gamma_F$ is finite and
positive, which means that the
HDP expands compressing the LDP.
In the particular case $p/q = 1$ and $\rho_{+} = 0$ the parameter
$\gamma_F$ appears to be a positive, logarithmically growing
with time function, which behavior  agrees
with the results of Ref.\cite{ara}.

(2) When 
$p (1 - \rho_{+}) <  
q ( 1 - \rho_{-})$ the parameter $\gamma_F$ is less than zero
- the expanding LDP and the applied force effectively
compress the HDP.

(3) When $p (1 - \rho_{+}) = 
q ( 1 - \rho_{-})$, the parameter $\gamma_F \equiv 0$.
This relation between the system parameters when 
the HDP and the LDP are in equilibrium with
each other and the TP mean displacement is zero, was found 
also in Ref.\cite{lebowitz2} from the analysis of the stationary behavior
in a finite 1D lattice gas.

"Shock front", or, in other words, the TP propagation in cases (1) and (2) has been 
discussed in detail in Ref.\cite{bur2}. Here, in view of our
interest 
in the Einstein relation, 
 we will focus on the case (3) in which situation 
the high- and low-density phases coexist such that the TP mean displacement is equal to zero.
On the other hand, it is clear that also in this case the TP will wander randomly
around the equilibrium position and one may expect that its mean-square displacement
will not be equal to zero.  

To determine the TP mean-square 
displacement $< X^2(t) >$
we may try to extend the approach outlined in Section 2 which will require some more cumbersome analysis. 
We will however pursue a different, heuristic approach and 
instead of evaluating $< X^2(t) >$ in terms of
the approach of Section 2 we will resort to 
the Einstein relation in Eq.(\ref{einstein}) assuming that it also holds in the inhomogeneous situation
under study. Our result will be then checked by numeric simulations.

To do this, we note first that contrary to the model of 
Section 3, here the TP does not move, on average, 
when the external force $F$ acting on this particle is not equal to zero,
but rather to some critical value $F_c$, such that
\begin{equation}
F_c = \beta^{-1} \ln\left(\frac{1 - \rho_-}{1 - \rho_+}\right).
\end{equation} 
Hence, it seems reasonable to define the 
TP mobility in this inhomogeneous case as
\begin{equation}
\mu(t) = \lim_{F \to F_c} \frac{V(t)}{(F - F_c)} = \frac{1}{\sqrt{t}} \lim_{F \to F_c} \frac{\gamma_F}{(F - F_c)}
\end{equation}
Further on, we find from Eq.(\ref{j}) 
that in the limit $\beta (F -  F_c) \ll 1$ the prefactor $\gamma_F$ follows
\begin{equation}
\gamma_F \sim \sqrt{\frac{2}{\pi}} \; \frac{p (1 - \rho_+) - q (1 - \rho_-)}{q \rho_- + p \rho_+},
\end{equation}
which implies that
$\mu(t)$
obeys
\begin{equation}
\mu(t) = \beta \frac{(1 - \rho_-)(1 - \rho_+)}{(\rho_- + \rho_+ - 2 \rho_- \rho_+)} \sqrt{\frac{2}{\pi t}}.
\end{equation}
Now, assuming that Eq.(\ref{einstein}) holds, we get
\begin{equation}
\label{harrr}
< X^{2}(t)>  \; = 
\; \frac{(1 - \rho_{-})(1 - \rho_{+})}{(\rho_{-} 
\; + \; \rho_{+} \; - \; 2 \; \rho_{-} \; \rho_{+})} 
\; \sqrt{\frac{8 t}{\pi}},
\end{equation}
which reduces to the classical result in Eq.(\ref{harr}) in the limit
$\rho_{-} = \rho_{+}$. 

\begin{figure}[h]
\begin{center}
\includegraphics*[scale=0.5]{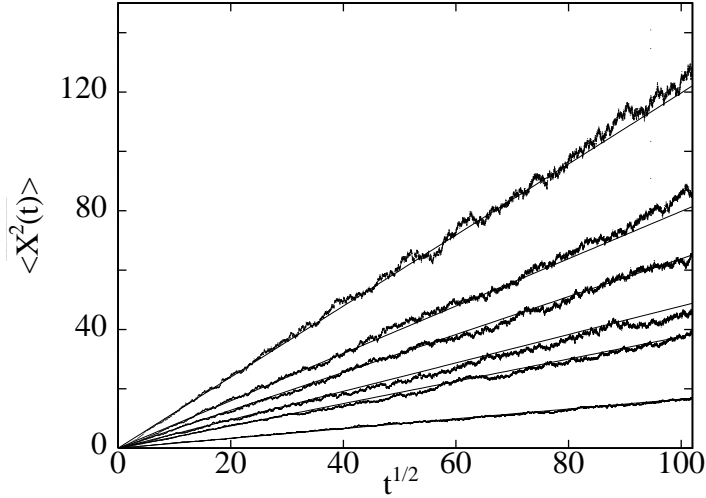}
\end{center}
\caption{\label{reseau}\small \sl Mean-square displacement of 
the TP in the critical case 
$p/q = (1 - \rho_{-})/(1 - \rho_{+})$. 
The solid lines show 
our analytical prediction from Eq.(\ref{harrr}) and 
 the noisy lines present the results of Monte Carlo simulations.
The curves from top to bottom correspond 
to the following values of the parameters 
$(\rho_{-},\rho_{+})$:
The first two curves are the 
analytical result in Eq.(\ref{harr}) 
and numerical data for the
symmetric cases  $(0.4,0.4)$ and $(0.5,0.5)$.  
The lower curves correspond to $(0.6,0.5)$,
$(0.9,0.4)$, $(0.7,0.5)$ and $(0.8,0.2)$
respectively.}
\end{figure}

In Fig.3 
we present results of extensive Monte Carlo simulations which 
confirm the 
result in Eq.(\ref{harrr})
and hence, signify that 
the Einstein relation seemingly works in this 
inhomogeneous case.

\section{Biased TP diffusion in a 1D lattice gas in contact with a reservoir.}

We turn now to discussion of the TP biased diffusion in situation in which the reservoir
is present and particle exchanges with the reservoir are allowed; that is, when both 
the parameters $f$ and $g$ are not equal to zero. 

In 1D systems with continuous 
particles exchanges with a reservoir, which situation is appropriate to 
adsorption on polymer chains \cite{1000,1100},
the biased TP will ultimately move with a constant velocity $V$ 
and particles distribution around the TP will be characterized
by stationary density profiles $k(\lambda)$
\cite{benichou}.    Here, the general solution
 of
 Eqs.(\ref{systemek1}) and (\ref{systemek2}) has the following form:
\begin{equation}
\label{dprofiles}
k(\lambda) =  \rho_s + K_{\pm} exp\Big(-\sigma |n|/\lambda_{\pm}\Big), 
\;\;\; \lambda = \sigma n, \;\;\; n \in Z,
\end{equation}
where the characteristic 
lengths $\lambda_{\pm}$ obey
\begin{equation}
\label{lambdas}
\lambda_{\pm}= \; \mp \; \sigma \; ln^{-1}\Big[
\frac{A_{1} + A_{-1} + 2 (f + g) \mp \sqrt{\Big(A_{1} + A_{-1} + 2 (f + g)\Big)^2 - 4 A_{1} A_{-1}}}{2 A_{1}}
\Big],
\end{equation} 
while the amplitudes $K_{\pm}$ are given respectively by
\begin{equation}
K_{+} = \rho_s \frac{A_{1} - A_{-1}}{A_{-1} - A_{1} \exp(- \sigma/\lambda_{+})}, \;\;\; \text{and} \;\;\; 
K_{-} = \rho_s \frac{A_{1} - A_{-1}}{A_{-1} \exp(- \sigma/\lambda_{-}) - A_{1} }. 
\end{equation}
Note that $\lambda_{-} > \lambda_{+}$, and consequently,  the local 
density past the TP approaches its non-perturbed value $\rho_s$
slower than in front of it;  this signifies that
 correlations between the TP position and particle
distribution are stronger past the TP. Next,  $K_{+}$ is always
 positive, while $K_{-} < 0$; this means that 
the density profile is a non-monotoneous function of $\lambda$ 
and is characterized by a jammed region in front
of the TP, in which the local density is 
higher than $\rho_s$, and a depleted region past the TP in which 
the density is lower than $\rho_s$. This is, of course, 
quite similar to the 
behavior observed in situations in which exchanges with the reservoir
are not allowed (Sections 3 and 4); essential difference is, however, 
that here the inhomogeneous 
particles distribution around the TP attains a stationary form.

\begin{figure}[h]
\begin{center}
\includegraphics*[scale=0.5]{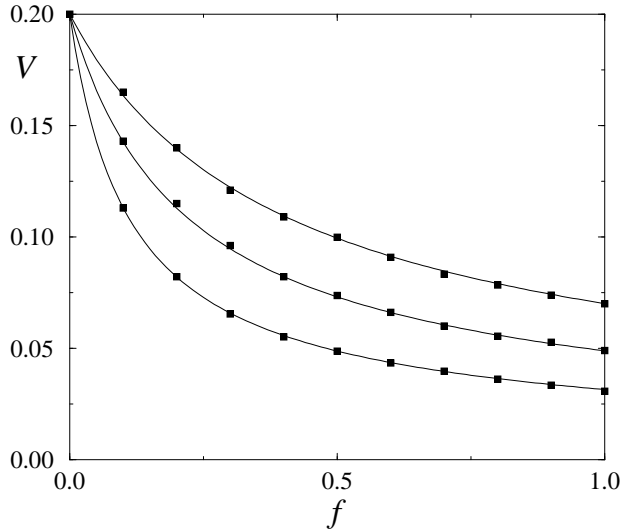}
\end{center}
\caption{\label{vitfig} \small \sl Terminal velocity of the TP 
as a function of the "creation" probability $f$
at different values of the parameter $g$. The TP 
hopping probabilities are $p_{+1} = 0.6$
and $p_{-1} = 0.4$. The solid lines  give the analytical solution
while the filled squares denote 
the results of Monte-Carlo simulations. Upper curves correspond to $g = 0.8$, the intermediate -
to $g = 0.5$ and the lower - to $g = 0.3$, respectively.}
\end{figure}  

\begin{figure}[h]
\begin{center}
\includegraphics*[scale=0.5]{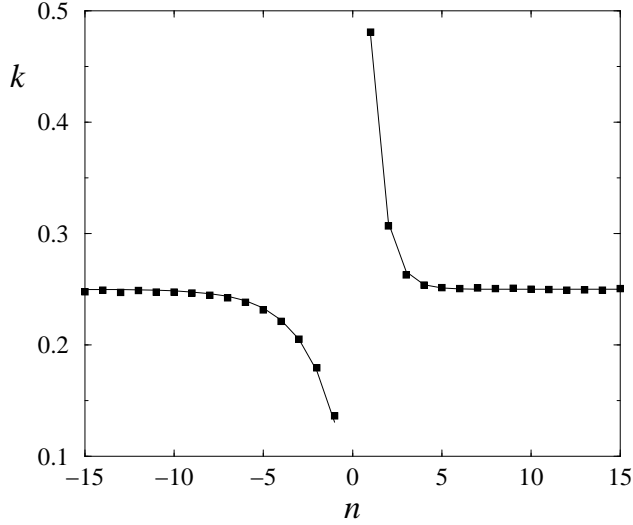}
\end{center}
\caption{\label{proffig} \small \sl Density profile around stationary moving TP for $f = 0.1$, $g = 0.3$ and $p = 0.98$. 
The solid line is the plot of the analytical solution.
Filled squares are the results of Monte-Carlo simulations.
}
\end{figure}

Now, we are in position to obtain a system of two closed-form non-linear equations determining implicitly the unknown 
parameters $A_{1}$ and $A_{-1}$, which will allow us to compute
 the TP terminal velocity, related to $A_{\pm 1}$ through 
$V = \sigma (A_{1} - A_{-1})/2 \tau^*$.
Substituting Eq.(\ref{dprofiles}) into Eq.(\ref{defA}), we find 
\begin{equation}
\label{o}
A_{1} = 1 + \frac{p_1 \tau^*}{\tau} \Big[1 - \rho_s - \rho_s \frac{A_{1} - A_{-1}}{A_{-1} exp(\sigma/\lambda'_{+}) - A_{1}}
\Big]
\end{equation}
and 
\begin{equation}
\label{b}
A_{-1} = 1 + \frac{p_{-1} \tau^*}{\tau} \Big[1 - \rho_s - \rho_s \frac{A_{1} - A_{-1}}{A_{-1}  - A_{1} exp(\sigma/\lambda'_{-})},
\Big]
\end{equation}
which system of equations defines the stationary 
velocity $V$ of the TP (see Fig.\ref{vitfig}) as well as the density profiles (see
Fig.\ref{proffig}). 
For arbitrary values of $p$, $f$ and $g$  the parameters $A_{\pm 1}$, defined by
Eqs.(\ref{o}) and (\ref{b}), and consequently, $V$ can be determined
only numerically (see Figs.4 to 6).  However, $V$ can be found analytically in the explicit form 
in the limit of
a vanishingly  small force $E$, $\beta E \ll 1$. 
Expanding $A_{\pm 1}$ in the Taylor series in powers of $E$ and retaining
only linear with $E$ terms, we find that
the TP velocity follows 
\begin{equation}
\label{st}
V \sim \zeta^{-1} E,
\end{equation}
which relation can be thought off as the analog of the Stokes formula for driven 
motion in a 1D lattice gas
undergoing continuous particles exchanges with the vapor phase. Eq.(\ref{st}) 
signifies that the frictional force exerted on the TP by the lattice gas
particles is $viscous$. The friction coefficient, i.e. the proportionality factor in 
Eq.(\ref{st}), is given explicitly by
\begin{equation}
\label{dfriction}
\zeta = \frac{2 \tau}{\beta \sigma^2 (1 - \rho_s)} \Big[1 + \frac{\rho_s \tau^*}{\tau (f + g)}
\frac{2}{1 + \sqrt{1 + 2 (1 + \tau^* (1 - \rho_s)/\tau)/(f + g)}} \Big]
\end{equation}
Note that $\zeta$ in Eq.(\ref{dfriction}) can be written down as the sum, $
\zeta=\zeta_{cm}+\zeta_{coop}$. The first term, $\zeta_{cm}=2\tau/ \beta \sigma^2(1-\rho_s)$
is a typical mean-field result and corresponds to a perfertly
homogeneous lattice gas (see discussion following Eq.(\ref{defk})). The second
term,
\begin{equation}
\zeta_{coop}=\frac{8\tau^*\rho_s}{\beta\sigma^2(1-\rho_s)(f+g)}\frac{1}{1+\sqrt{1 + 2 (1 + \tau^* (1 - \rho_s)/\tau)/(f + g)}},
\end{equation} 
has, however, a more complicated
origin. Namely,  it reflects a cooperative behavior
emerging in the lattice gas, associated with the formation of
inhomogeneous density profiles (see Fig.\ref{proffig}) - the formation of a ``traffic jam'' in front of
the TP and a ``depleted'' region  past the TP (for more details, see  Ref.\cite{benichou}).  The characteristic lengths of these two
regions as well as the amplitudes $K_{\pm}$ depend on the magnitude of the TP velocity; on the other hand, the TP velocity is itself
dependent on the density profiles, in virtue of Eq.(\ref{vitesse}). This results in an intricate interplay between the jamming
effect of the TP and smoothening of the created inhomogeneities by diffusive processes. Note also that cooperative behavior becomes
most prominent in  the conserved particle number limit \cite{bur2}. Setting $f,g \to 0$, while keeping their ratio fixed (which
insures that $\rho_s$ stays constant), one notices that $\zeta_{coop}$ gets infinitely large. As a matter of fact, as we have shown already in section 3, 
in such a situation no stationary density profiles around the TP exist.

In order to check
our analytical predictions, we have  performed numerical Monte
Carlo simulations. Results of these simulations, performed at different values
of the parameters $f$, $g$, and $p_1$, are also represented in
Figs.\ref{vitfig} and \ref{proffig}.

Consider finally the situation with $E = 0$, in which case the terminal velocity vanishes 
and one expects conventional diffusive motion 
with the mean square displacement of the form of Eq.(\ref{dif}). Heuristically,  we can evaluate  
$D$ 
for the system under study
if we assume the validity 
of the Einstein relation in Eq.(\ref{einstein}). Noticing that here the biased 
TP mobility $\mu$ is just $\mu = 1/\zeta$, we find \cite{benichou}:
\begin{equation}
\label{ddiffusion}
D = \frac{\sigma^2 (1 - \rho_s)}{2 \tau} \left\{1 +\frac{\rho_s\tau^*}{\tau(f+g)} 
\frac{2}{1 + \sqrt{1 +2 (1 + \tau^* (1 - \rho_s)/\tau) /(f+g)}} \right\}^{-1}.
\end{equation}
 Monte Carlo
simulations (see Fig.6) evidently
confirm our prediction for $D$ given by Eq.(\ref{ddiffusion}), and hence, confirm the 
validity of the Einstein relation for the system under study. This is, of course, not an unexpected, but still 
a non-trivial result \cite{lebowitz2}.

\begin{figure}[h]
\begin{center}
\includegraphics*[scale=0.5]{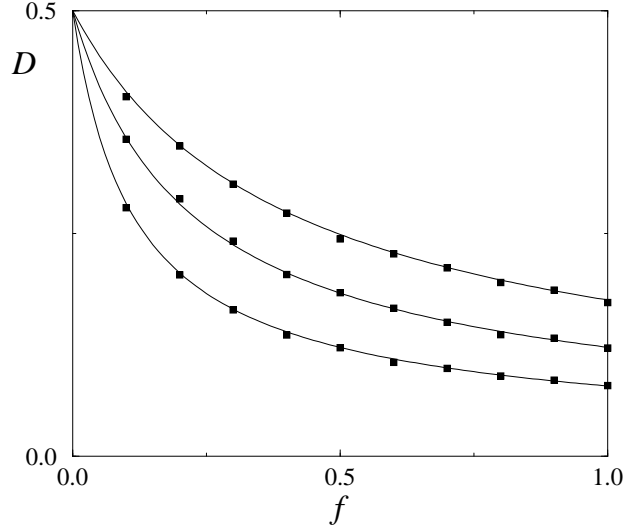}
\end{center}
\caption{\label{self}\small \sl 
Self-diffusion coefficient of the TP as a function of the adsorption probability $f$. 
Notations and values of
$g$ are the same as in Figs.\ref{vitfig}.}
\end{figure}

\section{Biased TP diffusion in 2D and 3D lattice gases.}

We turn now to the analysis of 
a biased TP diffusion in higher dimensional (2D and 3D) 
lattice gases 
exposed to a vapor phase. Here, the situation gets somewhat more difficult from the computational
 point of view;  we have now to
solve the partial difference equations problem (\ref{systemek1}),
(\ref{systemek2}) rather than the mere difference
equations arising in the 1D case.  

Below we briefly 
outline
such a solution for a 2D case. Results for three-dimensions are obtained 
along essentially the same lines and here we display them without derivation.
Solution in two-dimensions can be found in a most convenient fashion if we introduce 
the
generating function for the particle density profiles, defined as
\begin{equation}
H(w_1,w_2)\equiv\sum_{n_1=-\infty}^{+\infty}\sum_{n_2=-\infty}^{+\infty} \left(\rho_s - k(n_1,n_2)\right) w_1^{n_1}w_2^{n_2}
\label{defH}
\end{equation}
Multiplying both sides of Eqs. (\ref{systemek1}) and (\ref{systemek2}) by $w_1^{n_1}w_2^{n_2}$,
 and performing summations over $n_1$ and $n_2$,
we find that $H(w_1,w_2)$ is  given explicitly by
\begin{eqnarray}
H(w_1,w_2)&=& - \; K(w_1,w_2)\;\Big\{A_1w_{1}^{-1}+A_{-1}w_1+A_2(w_2+w_2^{-1})-\alpha\Big\}^{-1}, 
\label{adevelopper}
\end{eqnarray}
where
$
\alpha\equiv\sum_\nu A_\nu+4(f+g)$
and 
\begin{equation}
K(w_1,w_2)\equiv \sum_\nu
A_\nu(w_{|\nu|}^{\nu/|\nu|}-1) \left(k({\bf e}_{\boldsymbol
\nu})- \rho_s\right)+\rho_s(A_1-A_{-1})(w_{1}-w_1^{-1})
\label{defK}. 
\end{equation}
Equations (\ref{adevelopper}) and (\ref{defK}) determine the
generation function for the density profiles exactly.

Before we proceed to the inversion of $H(w_1,w_2)$ with respect to the variables $w_1$ and $w_2$, we note that
we can access 
interesting integral characteristics of the density profiles directly using the result in  Eqs.(\ref{adevelopper}) 
and (\ref{defK}). 
Namely, as we have already remarked,  the presence of the driven TP
 induces
 an inhomogeneous density distribution
in the lattice. One can thus pose a natural question whether 
equilibrium between adsorption and desorption processes gets shifted due to such a perturbancy, 
i.e. whether 
the equilibrium density in the lattice gas is different from that given by Eq.(\ref{rhostat}). The answer is trivially "no"
 in the case when the particles number is
explicitly conserved, but in the general case 
with arbitrary $f$ and $g$ this is not at all evident: similarly to the behavior in 1D system 
one expects that also in two-dimensions the density  profiles are asymmetric as seen from the stationary moving 
TP and are characterized by a
condensed, "traffic-jam"-like region in front of and a depleted region past the TP. One anticipates then that the
desorption events are 
favored in front of the TP, while  the adsorption events are evidently suppressed by the excess density.
On the other hand, past the TP desorption is diminished due to 
the particles depletion while adsorption
may proceed more readily due to the same reason.
 It is thus not at all clear ${\it a \; priori}$ whether these two
effects can compensate each other exactly, in view of a possible
asymmetry of the density profiles, as it happens in the
1D model (see Fig.\ref{proffig}).

For this purpose, we study the behavior of  the integral
deviation $\Omega$ of the density from the equilibrium value $\rho_{s}$, i.e.
$\Omega\equiv\sum_{n_1=-\infty}^{+\infty}\sum_{n_2=-\infty}^{+\infty}\Big(\rho_s - k(n_1,n_2)\Big)$,
which can be computed straightforwardly from
 Eqs.(\ref{adevelopper}) and  (\ref{defK}) by setting both $w_1$ and $w_2$ equal to unity.  Noticing that
$K(w_1=1,w_2=1)=0$, and that
$A_1+A_{-1}+2A_2-\alpha=-4(f+g)$,  we obtain then
that $\Omega$ is strictly equal to 0.  This implies, in turn,  that the 
perturbancy of the density distribution in the lattice gas created by the driven TP
does not shift the global balance between the adsorption and desorption events. 

Inversion of the generating function with respect to $w_1$ and $w_2$ 
requires quite an involved mathematical analysis, which has been presented in detail in
 Ref.\cite{benichouPRB}. General solution
for the density profiles reads:
\begin{eqnarray}
k(n_1,n_2)=\rho_s &+& 
\alpha^{-1}\Big\{\sum_\nu A_\nu \; \Big(k({\bf e}_{\boldsymbol \nu}) - \rho_s\Big) \;  \nabla_{-\nu}F_{n_1,n_2}- \nonumber\\
&-&\rho_s(A_1-A_{-1})(\nabla_1-\nabla_{-1})F_{n_1,n_2}\Big\}.
\label{2dsolh}
\end{eqnarray}
with
\begin{eqnarray}
F_{n_1,n_2}=
\left(\frac{A_{-1}}{A_1}\right)^{n_1/2}\int_0^{\infty}e^{-t}{\rm I}_{n_1}\left(2\alpha^{-1}\sqrt{A_1A_{-1}}t\right){\rm I}_{n_2}\left(2\alpha^{-1}A_2t\right){\rm d}t,
\label{repint}
\end{eqnarray}
where ${\rm I}_n(z)$ stands for the  modified Bessel function. We mention that
$F_{n_1,n_2}$  has   an interesting physical interpretation in terms of
the generating function of a random walk of a single particle (that
is, in absence of the particles environment) \cite{hughes}.

Now, the Eqs.(\ref{2dsolh}) and (\ref{repint}) display $k(n_1,n_2)$
as a function of the coefficients $A_\nu$ that remain to be determined.
As a matter of fact, these coefficients depend  themselves
on the local densities in the immediate vicinity of the tracer, i.e. on $k({\bf e_\nu})$. 
This implies that we have to determine them  from Eqs.(\ref{2dsolh}) 
and (\ref{repint}) in a self-consistent way \cite{benichouPRB}. Some analysis (see Ref.\cite{benichouPRB}) shows that $A_\nu$ are
determined implicitly as the solution of the following system of three non-linear matrix equations
\begin{equation}
\forall\nu=\{\pm1,2\},\;\;\;\;A_\nu=1+\frac{4\tau^*}{\tau}p_\nu\left\{1-\rho_s-\rho_s(A_1-A_{-1})\frac{\det\tilde{C}_\nu}{\det\tilde{C}}\right\},
\label{2dimplicite}
\end{equation}  
where
\begin{equation}
\tilde{C}=
\begin{pmatrix}
A_1\nabla_{-1}F_{\bf e_1}-\alpha & A_{-1}\nabla_1F_{\bf e_1} & A_{2}\nabla_{-2}F_{\bf e_1}\\
A_1\nabla_{-1}F_{\bf e_{-1}} & A_{-1}\nabla_1F_{\bf e_{-1}}-\alpha & A_{2}\nabla_{-2}F_{\bf e_{-1}}\\
A_1\nabla_{-1}F_{\bf e_2} & A_{-1}\nabla_1F_{\bf e_2} &  A_{2}\nabla_{-2}F_{\bf e_2}-\alpha
\end{pmatrix},
\end{equation}
the matrix  $\tilde{C_\nu}$ stands for the matrix obtained from  $\tilde{C}$
by replacing the $\nu$-th column  by the column-vector
$\tilde{F}$,
\begin{equation}
\tilde{F}=
\begin{pmatrix}
(\nabla_1-\nabla_{-1})F_{\bf e_1}\\
(\nabla_1-\nabla_{-1})F_{\bf e_{-1}}\\
(\nabla_1-\nabla_{-1})F_{\bf e_2}
\end{pmatrix},
\end{equation}
while $k({\bf e}_{\boldsymbol \nu})$ are expressed in terms of $A_\nu$ as
\begin{equation}
k({\bf e}_{\boldsymbol \nu})=1+\frac{\tau}{4\tau^*p_\nu}(1-A_\nu).
\label{ppasnul}
\end{equation}  
Lastly, we find that in 2D the TP terminal velocity obeys
\begin{equation}
V = \frac{\sigma}{\tau}(p_1-p_{-1})(1-
\rho_s)\Big\{1+\rho_s\frac{4\tau^*}{\tau}\frac{p_1\det\tilde{C}_1-p_{-1}\det\tilde{C}_{-1}}{\det\tilde{C}}\Big\}^{-1},
\label{forcevitesse}
\end{equation}
which represents the 
general force-velocity relation 
for the system under study,  valid for arbitrary
magnitude of the external bias and 
arbitrary values of 
other system's parameters, except
apparently the situation when  
$\tau^* =  \infty$, $f,g = 0$ and their ratio $f/g = const$. This case 
corresponds 
to the percolation-type problem describing biased diffusion 
in a randomly diluted lattice - the model which 
has been discussed in the Introduction. 
Some analysis shows that $V$ defined by Eq.(\ref{forcevitesse}) does not vanish 
for any value of external bias $E$. This signifies that the 
decoupling approximation underlying the derivation of
the result in Eq.(\ref{forcevitesse}) is, of course, 
inappropriate for situations with $quenched$ disorder, but only for situations
in which some mixing, either due to exchanges with the reservoir, or due to diffusional processes,
is present.

We turn now to the limit $\beta E \ll 1$, in which case the problem simplifies considerably and allows to obtain explicit results
for the local densities in the immediate vicinity of the TP and consequently, for the TP terminal velocity and
its diffusivity.

In this limit, we arrive again at a Stokes-type formula of the form $V \sim E/ \zeta $, where now
\begin{equation}
\label{rrr}
\zeta=\frac{4\tau}{\beta\sigma^2(1-\rho_s)}\left\{1
+\frac{\tau^*}{\tau}\frac{\rho_s}{\Big(f + g + 1 + \tau^* (1 - \rho_s)/\tau\Big)\Big({\cal L}_2(x)-x\Big)}\right\},
\end{equation} 
with
\begin{equation}
x = \frac{1}{2} \frac{1 + \tau^* (1 - \rho_s)/\tau}{f + g + 1 + \tau^* (1 - \rho_s)/\tau}, \;\;\; \text{and} \;\;\;
 {\cal L}_2(x)\equiv\left\{\int_0^\infty e^{-t}\Big(({\rm I}_0(xt)-{\rm I}_2(xt)){\rm I}_0(xt){\rm d}t\right\}^{-1}.
\end{equation}
Note  that we again are able to single out two physically meaningful contributions to 
the friction coefficient $\zeta$.  Namely, the first term on the rhs of 
Eq.(\ref{rrr}) is just the mean-field-type result
corresponding to a perfectly stirred lattice gas, in which 
correlations between the TP and the lattice gas particles are discarded.
The second term mirrors the cooperative behavior emerging in the lattice gas
and is associated with the backflow effects.  In contrast to the 1D case, however, the contribution to the overall
friction coefficient stemming out of the cooperative effects 
remains finite in the conserved particles number limit.

Lastly, assuming {\it a priori} that the Einstein relation holds for the system
under study, 
we find the TP diffusion coefficient: 
\begin{equation}
D=\frac{\sigma^2}{4\tau}(1-\rho_s)\left\{1-\frac{2\rho_s\tau^*}{\tau}\frac{1}{ 4 
(f + g + 1 + \tau^* (1 - \rho_s)/\tau) {\cal L}_2(x)-1+(3\rho_s-1)\tau^*/\tau}\right\}.
\label{2dautodiffgen}
\end{equation}
It seems now interesting to compare our general result in Eq.(\ref{2dautodiffgen}) against the classical result of Nakazato and
Kitahara \cite{nakazato}, which describes TP
 diffusion coefficient in a 2D lattice gas with conserved
particles number. 
Setting $f$ and $g$ equal to zero, while assuming that their ratio 
has a  fixed value, $f/g=\rho_s/(1-\rho_s)$, we find
that the right-hand-side of Eq.(\ref{2dautodiffgen}) attains the form
\begin{equation}
\label{comparNaka}
\hat{D}=\frac{\sigma^2}{4\tau}(1-\rho_s)\left\{1-
\frac{2\rho_s\tau^*}{\tau}\frac{1-2/\pi}{1+(1-\rho_s)\tau^*/\tau-(1-2/\pi)(1+(1-3\rho_s)\tau^*/\tau)}\right\},
\end{equation}
which expression
 coincides with the earlier result obtained  in 
Refs.\cite{nakazato} and \cite{tahir} within the framework of a different, compared to ours,
 analytical technique. The result in Eq.(\ref{comparNaka}) is
known to be exact in the limits $\rho_s \ll 1$ and $\rho_s \sim 1$, and serves as a very good approximation for the
self-diffusion coefficient in hard-core lattice gases of arbitrary density \cite{kehr}.

In a similar fashion, 
we find that in 3D the TP terminal velocity obeys $V \sim E/ \zeta $, 
where 
\begin{equation}
\zeta = \frac{ 6 \tau}{\beta \sigma^2 (1 - \rho_s)} + \frac{12 \rho_s \tau^*}{\beta \sigma^2 l (1-\rho_s).
\Big(\alpha_0{\cal L}_3(2A_0/ \alpha_0)-A_0\Big)}, 
\end{equation}
with
\begin{equation}
A_0 = 1 + \frac{\tau^{*}}{l \tau} (1 - \rho_s), \;\;\; \alpha_0 = 6 \Big(1 + \frac{\tau^{*}(1 - \rho_s)}{l \tau} + \frac{f+g}{l}\Big),
\end{equation}
and 
\begin{equation}
{\cal L}_3(x)\equiv\left\{\int_0^\infty e^{-t}{\rm I}_0^{2}(xt)\left({\rm I}_0(xt)-{\rm I}_2(xt)\right){\rm d}t\right\}^{-1} = 
\left\{P({\bf 0};3x)-P(2{\bf e_1};3x)\right\}^{-1},
  \end{equation}
$P({\bf R};\xi)$ being the generating function,
  \begin{equation}
  P({\bf R};\xi)\equiv\sum_{j=0}^{+\infty}P_j({\bf R})\xi^j,
  \end{equation}
of the probability $P_j({\bf R})$ that a walker starting at the origin and performing a Polya random walk on
the sites of a 3D cubic lattice will arrive on the $j$-th step to
the site with the lattice vector  ${\bf R}$ \cite{hughes}.

Assuming next that 
the Einstein relation holds, we find 
that in 3D the TP
diffusion coefficient reads:
\begin{equation}
  D=\frac{\sigma^2(1-\rho_s)}{6\tau}\left\{1-\frac{2\rho_s\tau^*}{l\tau}\Big(\alpha_0{\cal
L}_3(2A_0/\alpha_0)-1+\frac{\tau^* (3\rho_s-1)}{l\tau}\Big)^{-1}\right\}
  \end{equation}
In the particular case of  conserved particles number, when $f,g \to 0$ but their ratio $f/g$ is kept fixed, $f/g =
\rho_s/(1-\rho_s)$, the latter equation reduces to
  \begin{equation}
\label{nk}
\hat{D}=\frac{\sigma^2(1-\rho_s)}{6\tau}\left\{1-\frac{2\rho_s\tau^*}{l\tau} \Big(6 A_0{\cal
L}(1/3)-1+\frac{\tau^*  (3\rho_s-1)}{l\tau}\Big)^{-1}\right\},
  \end{equation}
which coincides again 
with the classic
result 
obtained earlier in Refs.\cite{nakazato} and \cite{tahir}.

Asymptotical behavior of the density profiles at large distances from
the TP follows from the analysis of the analyticity properties of the complex function
$N(z)\equiv\sum_{n=-\infty}^{+\infty} k(n,0) z^n$. It has been  shown in  Ref.\cite{benichouPRB}
that in front of the TP,
$k(n,0)$ always decays exponentially with the distance:
\begin{equation}
k(n,0) \sim \rho_s + K_+\frac{\exp\Big(-n/ \lambda_+\Big)}{n^{1/2}},
\end{equation}
where the characteristic length $\lambda_+$ obeys, in two-dimensions:
\begin{equation}
\lambda_+ =
\ln^{-1}\Big(\frac{1}{A_{-1}}\left\{\frac{\alpha}{2}-A_2+\sqrt{\left(\frac{\alpha}{2}-A_2\right)^2-A_1A_{-1}}\right\}\Big)
\end{equation}
Note that $\lambda_+$ stays finite for any values of the system parameters.
On contrary, the behavior of the density profiles at large
distances past the tracer qualitatively  depends on the physical
situation studied. In the general case when  exchanges with the
particles reservoir are allowed, the decay of the density
profiles is still exponential:
\begin{equation}
k(-n,0) \sim \rho_s +  K_-\frac{\exp\Big(-n/ \lambda_-\Big)}{n^{1/2}},
\end{equation}    
where
\begin{equation}
\lambda_- = -
\ln^{-1}\Big(\frac{1}{A_{-1}}\left\{\frac{\alpha}{2}-A_2-\sqrt{\left(\frac{\alpha}{2}-A_2\right)^2-A_1A_{-1}}\right\}\Big)
\end{equation}
Note that in the general case the characteristic lenghts again, similarly to the 1D case,
satisfy the inequality $\lambda_->\lambda_+$, which means that the
correlations between the TP  and the lattice particles  are
always stronger past the TP than in front of it.

Such correlations  become extremely strong in the special case when the particles
exchanges with the vapor phase are forbidden, i.e., in the conserved particles number limit. 
In this case, we have that $\lambda_-$ becomes infinitely large 
and, in the limit $n\to+\infty$, 
the particle density follows
\begin{equation}
k(-n,0) = \rho_s -\frac{K_{-}'(d = 2)}{n^{3/2}}\left(1+\frac{3}{8n}+{\mathcal O}\Big(\frac{1}{n^2}\Big)\right), 
\;\;\; K_-'(d = 2) > 0.
\label{algebrique}
\end{equation} 
Remarkably enough, in this case the correlations between the TP position
and the particles distribution
vanish  {\it algebraically} slow with the distance! This implies, in turn,
 that in the conserved particles
number case, the diffusive 
mixing of the lattice gas is not efficient enough to prevent the appearence of the quasi-long-range order
and the medium "remembers"
 the passage of the TP
on a long time and space scale, which signifies very 
strong memory effects.
In three dimensions, an analogous result for the asymptotical behavior of the density profiles
as seen from the stationary moving TP is:
\begin{equation}
k(- n,0,0) \sim \rho_s   - \frac{K_{-}'(d = 3) \ln(n)}{n^2}, \;\;\; K_{-}'(d = 3) > 0,
\end{equation}
where $K_{-}'(d = 3)$ is a constant \cite{25}.  

\section{Biased, single vacancy-mediated TP diffusion.}

We turn finally to the extreme case of a very dense lattice gas in which
all sites except one (called in what follows "a vacancy") are 
filled with identical neutral, hard-core particles (see, Fig.7). 
The particles move randomly 
by exchanging their positions with this single vacancy, subject to 
the hard-core exclusion constraint. 
The just described model, which represents, in fact, one of the simplest cases
of the so-called "slaved
diffusion processes", 
has been studied  over the years in various guises, such as, for instance,
the "constrained dynamics" model of Palmer \cite{palmer}. 

\begin{figure}[h]
\begin{center}
\includegraphics*[scale=0.3]{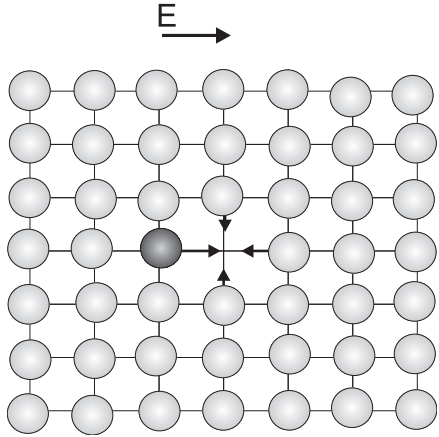}
\end{center}
\caption{\label{self} \small \sl Two-dimensional, infinite in both directions,
square lattice in which all sites except one are filled with identical hard-core
particles (gray spheres). The black sphere denotes a single tracer particle, which is
subject to external field $\bf{E}$, oriented in the positive $x_1$ direction, and
thus has asymmetric hopping probabilities. The arrows
of different size depict schematically the hopping probabilities; a larger arrow
 near the TP indicates that it has a preference for 
moving in the direction of the applied field.}
\end{figure}  
 
Brummelhuis and Hilhorst \cite{hilhorst} were first to
present an exact solution of this model in the
lattice formulation. It has been shown that in the presence of a
single vacancy the TP trajectories are remarkably confined;   the
mean-square displacement shows an unbounded growth, but it  
grows only $logarithmically$ with time,
\begin{equation}
\label{log}
<{\bf R}_n^2> \sim \frac{\ln(n)}{\pi (\pi - 1)}, \;\;\; \text{as $n \to\infty$},
\end{equation}     
which implies that the TP diffusivity $D_n$ obeys
\begin{equation}
\label{Dii}
D_n \sim \frac{\ln(n)}{4 \pi (\pi - 1) n}.
\end{equation}
Brummelhuis and Hilhorst have also found \cite{hilhorst} 
that at sufficiently large times the probability  $P_n({\bf R}_n)$  of finding the TP at time moment $n$ at position
${\bf R}_n$  
converges to a limiting form as a
function of the scaling variable $\eta =  |{\bf R}_n|/\sqrt{\ln(n)}$. Still striking,
this limiting distribution is not a $Gaussian$ but a modified Bessel
function $K_0(\eta)$, which signifies that the successive steps of the TP, although
separated by long time intervals, are effectively correlated. These results have been
subsequently reproduced by means of different analytical techniques in 
Refs.\cite{newman,toro} and \cite{zia}.

As in the previous sections, we focus here on TP 
dynamics in the biased case, when the TP 
is charged, while the lattice gas particles are neutral, 
and the system is subject to a 
constant electric field. We aim to determine exactly 
the TP mean-displacement $< {\bf R}_n >$, 
which will give us an access to the TP 
mobility and thus will allow to verify the validity 
of the Einstein relation in this model with remarkably confined
dynamics.

A standard
 approach to define the properties of the TP random 
walk, as exposed in Section 2,  would be to start with a master 
equation determining the evolution of the whole
configuration of particles. In doing so, 
similarly to the analysis of the tracer diffusion on 
2D lattices in the presence of a finite vacancy
concentration (see, e.g., Ref.\cite{benichou}), 
one obtains the evolution of the joint distribution  
$P_n({\bf R}_n,{\bf Y})$ of the TP position ${\bf R}_n$ and of the
vacancy position ${\bf Y}$ at time moment $n$. 
The property of interest, i.e. the reduced distribution function 
of 
the TP alone  will then be found from 
$P_n({\bf R}_n,{\bf Y})$ by performing lattice summation over 
all possible values of the variable ${\bf Y}$.

Here we pursue, however, a different approach, which has 
been first put forward in the original work  of  
Brummelhuis and Hilhorst \cite{hilhorst}; that is,  we   
construct the distribution function of the TP position at time $n$  directly
in terms of the return probabilities of the random walk performed by the vacancy. 
The only complication, compared to the unbiased case considered by
Brummelhuis and Hilhorst \cite{hilhorst}, is that in our case  ten  different 
return probabilities will be
involved, instead of three different ones appearing  in the unbiased case. 
Hence, the analysis will be slightly more involved.

We begin by introducing some basic
notations. Let
\begin{itemize}
\item $P_n({\bf R}_n)$ be the probability that the TP, which starts its random walk at the origin, appears 
at the site ${\bf R}_n$ at time moment $n$, given that the vacancy is initially at site ${\bf Y}_0$.
\item $F^*_n({\bf 0}\;|\;{\bf Y_0})$ be the probability that the
vacancy, which starts its random walk at the site ${\bf Y}_0$,
 arrives at the origin  ${\bf 0}$ for the first time at the time step $n$.
\item $F^*_n({\bf 0}\;|\;{\bf e}_{\boldsymbol \nu}\;|\;{\bf Y_0})$ be the
conditional probability that the vacancy, which starts its random walk at the site ${\bf Y}_0$, 
 appears  at the origin for the first
time at the time step $n$, being at time moment $n - 1$ at the site 
${\bf e}_{\boldsymbol \nu}$. 
\end{itemize}
Further on, for any time-dependent quantity $L_n$ we define the
generating function:
\begin{eqnarray}
\label{L}
L(\xi)=\sum_{n=0}^{+\infty}L_n \xi^n,
\end{eqnarray}
and for any space-dependent quantity $Z({\bf X})$ the discrete Fourier
transform:
\begin{eqnarray}
\label{Y}
{\widetilde Z}({\bf k})=\sum_{\bf X} \exp\Big( i ({\bf k} \cdot {\bf X}) \Big) Y({\bf X}),
\end{eqnarray}
where the sum with the subscript ${\bf X} = (x_1,x_2)$ 
runs over all lattice sites.

Now, following Brummelhuis and Hilhorst \cite{hilhorst}, we write down directly 
the equation obeyed by the reduced probability distribution 
$P_n({\bf R}_n)$ (cf Ref.\cite{toro} for a study of the joint
probability of the TP position and of the vacancy position in the unbiased case):
\begin{eqnarray}
P_n({\bf R}_n)&=& \delta_{{\bf R}_n,{\bf 0}}\left(1-\sum_{j=0}^n F^*_j({\bf
0}|{\bf Y_0})\right) + \nonumber\\ 
&+&\sum_{p=1}^{+\infty}\sum_{m_1=1}^{+\infty}\ldots\sum_{m_p=1}^{+\infty}
\sum_{m_{p+1}=0}^{+\infty}\delta_{m_1+\ldots+m_{p+1},n}\sum_{\nu_1}\ldots\sum_{\nu_p}\delta_{{\bf
e}_{{\boldsymbol \nu}_1}+\ldots+{\bf
e}_{{\boldsymbol \nu}_p},{\bf R}_n} \times \nonumber\\
&\times& \left(1-\sum_{j=0}^{m_{p+1}}F^*_j({\bf
0}|-{\bf e}_{{\boldsymbol \nu}_{\bf p}})\right) \times \nonumber\\
&\times& F^*_{m_p}({\bf 0}|{\bf e}_{{\boldsymbol \nu}_{\bf p}}\;|\;-{\bf
e}_{{\boldsymbol \nu}_{\bf p-1}})\ldots F^*_{m_2}({\bf 0}|{\bf e}_{{\boldsymbol
\nu}_{\bf 2}}\;|\;-{\bf e}_{{\boldsymbol \nu}_{\bf 1}}) F^*_{m_1}({\bf 0}|{\bf
e}_{{\boldsymbol \nu}_{\bf 1}}\;|\;{\bf Y_0}).
\label{Ptr}
\end{eqnarray}
Next, using the definition of the generating functions and of the discrete Fourier transforms, Eqs.(\ref{L}) and (\ref{Y}), 
we obtain 
the following matricial representation of the generating function of
 the TP probability distribution:
\begin{equation}
\label{P}
\widetilde{P}({\bf k};\xi)=\frac{1}{1-\xi}\left(1+{\cal D}^{-1}({\bf
k};\xi)\sum_{\mu}U_{ \mu}({\bf k};\xi)F^*({\bf
0}\;|\;{\bf e}_{\boldsymbol \mu}\;|\;{\bf Y_0};\xi)\right).
\end{equation}
In Eq.(\ref{P}) 
the function ${\cal D}({\bf k};\xi)$ stands for 
the determinant of the following $4 \times 4$ matrix,
\begin{equation}
\label{D}
{\cal D}({\bf k};\xi)\equiv{\rm det}({\rm{\bf  I-T}}({\bf k};\xi)),
\end{equation} 
where ${\rm {\bf T}}({\bf k};\xi)$ is a $4 \times 4$ matrix of the form: 
\begin{equation}
{\rm {\bf T}}({\bf k};\xi)\equiv
\begin{pmatrix}
e^{ik_1} A_{1,-1}(\xi) & e^{ik_1} A_{1,1}(\xi) &  e^{ik_1} A_{1,-2}(\xi) &  e^{ik_1} A_{1,2}(\xi) \\
e^{-ik_1} A_{-1,-1}(\xi) & e^{-ik_1} A_{-1,1}(\xi) & e^{-ik_1} A_{-1,-2}(\xi) & e^{-ik_1} A_{-1,2}(\xi) \\
e^{ik_2} A_{2,-1}(\xi) & e^{ik_2} A_{2,1}(\xi) & e^{ik_2} A_{2,-2}(\xi) & e^{ik_2} A_{2,2}(\xi) \\
e^{-ik_2} A_{-2,-1}(\xi) & e^{-ik_2} A_{-2,1}(\xi) & e^{-ik_2} A_{-2,-2}(\xi) & e^{-ik_2} A_{-2,2}(\xi) 
\end{pmatrix},
\end{equation}
and 
the coefficients $A_{\nu,\mu}(\xi)$, $\nu,\mu = \pm 1, \pm 2$,  defined as
\begin{equation}
A_{\nu,\mu}(\xi) \equiv F^*({\bf 0}\;|\;{\bf e}_{\boldsymbol \nu}\;|\;{\bf e}_{\boldsymbol \mu};\xi) = \sum_{n = 0}^{+ \infty}  
F^*_n({\bf 0}\;|\;{\bf e}_{ \boldsymbol \nu}\;|\;{\bf e}_{\boldsymbol \mu}) \xi^n, 
\end{equation}
are the generating functions of the conditional probabilities for the 
first time visit of the origin
by the vacancy. 
Lastly, the matrix $U_{{\boldsymbol \mu}}({\bf k};\xi)$ in Eq.(\ref{P}) is given by
\begin{eqnarray}
U_{\mu}({\bf k};\xi)\equiv{\cal D}({\bf k};\xi)\sum_{ \nu} 
(1-e^{-i ({\bf k} \cdot {\bf e}_{\boldsymbol \nu})})({\rm I} -{\rm T}({\bf
k};\xi))^{-1}_{\nu,\mu} \; e^{i ({\bf
k} \cdot {\bf e}_{\boldsymbol \mu})}.
\end{eqnarray}
The property of interest - the TP probability distribution function,  will be then obtained  by inverting
$\widetilde{P}({\bf k};\xi)$ with respect to the wave-vector $k$ and to the variable $\xi$:
\begin{eqnarray}
P_n({\bf R}_n)=\frac{1}{2i\pi}\oint _{\cal C} \frac{{\rm
d}\xi}{\xi^{n+1}}\frac{1}{(2\pi)^2}\int_{-\pi}^{\pi}{\rm
d}k_1\int_{-\pi}^{\pi}{\rm d}k_2\;e^{i ({\bf k} \cdot {\bf R}_n)}\widetilde{P}({\bf k};\xi),
\end{eqnarray}
where the contour of integration ${\cal C}$ encircles the origin counterclockwise.

After straightforward but rather 
tedious calculations (see, Ref.\cite{olli}), we find that
in the small-${\bf k}$ limit and $\xi \to 1^-$,  which defines the leading large-$n$ behavior  of the
probability distribution,
the generating function $\widetilde{P}({\bf k};\xi)$ obeys
\begin{eqnarray}
\label{V}
\widetilde{P}({\bf
k};\xi)=\frac{1}{1-\xi}\Big\{1-\left(-i \alpha_0 k_1 +
\frac{1}{2} \alpha_1 k_1^2+\frac{1}{2}\alpha_2 k_2^2\right)\ln{(1-\xi)}\Big\}^{-1},
\end{eqnarray}
where the coefficients
\begin{equation}
\label{alpha}
\begin{cases}
\alpha_0(E)\equiv \pi^{-1} \sinh(\beta E/2)((2\pi-3)\cosh(\beta
E/2)+1)^{-1},\\
\alpha_1(E)\equiv \pi^{-1} \cosh(\beta E/2)((2\pi-3)\cosh(\beta
E/2)+1)^{-1},\\
\alpha_2(E)\equiv \pi^{-1} (\cosh(\beta
E/2)+2\pi-3)^{-1},
\end{cases}
\end{equation} 
are all functions of the field strength $E$ and of the temperature only. 

Now, the leading large-$n$ asymptotical behavior of the TP mean displacement can be obtained directly 
from Eq.(\ref{V}), since
the generating function of
the TP mean displacement:
\begin{equation} 
<{\bf R}(\xi)> \equiv \sum_{n =0}^{+\infty} < {\bf R}_n > \xi^n,
\end{equation}
obeys (see, e.g., Ref.\cite{kehr}):  
\begin{eqnarray}
\label{LL}
<{\bf R}(\xi)> =-i\left(\,\frac{\partial\widetilde{P} }{\partial k_1} ({\bf
0};\xi) \,{\bf e_1}\,+\,\frac{\partial\widetilde{P} }{\partial k_2} ({\bf
0};\xi) \,{\bf e_2}\,\right).
\end{eqnarray}
Consequently, 
differentiating  the expression on the right-hand-side side of  Eq.(\ref{V}) with respect to the components of
the
wave-vector ${\bf k}$, 
we find that the asymptotical behavior of the generating function of
the TP mean displacement in the vicinity of $\xi = 1^-$ follows
\begin{eqnarray}
<{\bf R}(\xi)> \sim \Big(\frac{\alpha_0(E)}{1-\xi}\,\ln{\frac{1}{1-\xi}}\Big) {\bf e_1}.
\end{eqnarray}
Further on, using the discrete Tauberian theorem
(cf. Ref.\cite{hughes}) and Eq.(\ref{alpha}), we find the following 
force-velocity relation for the system under study
\begin{eqnarray}
<{\bf R}_n > \sim  \Big(\alpha_0(E) \,\ln{n}\Big) \;{\bf e_1} =  \Big(\frac{1}{\pi} \frac{\sinh(\beta
E/2)}{(2\pi-3)\cosh(\beta
E/2)+1} \,\ln{n}\Big) \;{\bf e_1}, \;\;\; \text{as $n \to \infty$},
\label{vitessee}
\end{eqnarray}
which shows that the TP mean displacement 
grows $logarithmically$ with $n$.  In the limit $\beta E \ll 1$, 
 the coefficient $\alpha_0(E)$ obeys
\begin{equation}
\alpha_0(E) = \frac{\beta E}{4 \pi (\pi - 1)} + {\mathcal O}(E^3),
\end{equation}
and hence, the mobility $\mu_n$, defined in Eq.(\ref{mob}), follows
\begin{equation}
\label{mob2}
\mu_n \sim \frac{\beta}{4 \pi (\pi - 1)} \; \frac{\ln{(n)}}{n}, \;\;\; \text{as $n \to \infty$}.
\end{equation}
Comparing next the result in Eq.(\ref{mob2}) 
with that for the diffusivity $D_n$, Eq.(\ref{Dii}), 
derived by Brummelhuis and
Hilhorst \cite{hilhorst} for 
the unbiased case, we infer that the TP mobility and diffusivity do obey, at least in the leading
in $n$ order, the generalized Einstein relation 
in Eq.(\ref{einstein})!
 Note, that this can not be, 
of course,
an  $\it a \; priori$ expected result, 
in view of an intricate nature of the random walks involved and anomalous, 
$logarithmic$ confinement of the random walk trajectories.

We turn next to calculation of the asymptotic 
forms of the probability distribution $P_n({\bf R}_n)$. 
Inverting 
$\widetilde{P}({\bf
k};\xi)$ with respect to ${\bf k}$, we
find that in the 
large-$n$ and large-$X$ limits, in the general case when $ 0 \geq E < \infty$, $P_n({\bf R}_n)$ obeys
\begin{eqnarray}
\label{U}
P_n({\bf
R}_n)\sim \Big(\pi \sqrt{\alpha_1(E) \alpha_2(E)}
 \ln{(n)} \Big)^{-1} \; \exp\Big(\frac{\alpha_0(E)}{\alpha_1(E)} x_1\Big) \; K_0(\eta_E(n)),
\end{eqnarray}
where $K_0$ is the modified Bessel (McDonald) function of 
zeroth order, while
\begin{equation}
\label{KK}
\eta_E(\lambda) \equiv \sqrt{\frac{2}{\ln{(\lambda)}}+\frac{\alpha_0^2(E)}{\alpha_1(E)}}
\sqrt{\frac{x_1^2}{\alpha_1(E)}+\frac{x_2^2}{\alpha_2(E)}},
\end{equation}
$x_1$ and $x_2$ being the components of the vector ${\bf R}_n$. The distribution in Eq.(\ref{U})
reduces to the result obtained by Brummelhuis and Hilhorst \cite{hilhorst} in the limit $E = 0$.

\section{Conclusions.}

To conclude, in 
this presentation we have 
overviewed
some recent results on 
biased tracer diffusion in
disordered dynamical
environments 
represented by
hard-core lattice gases.
We have considered several possible situations
including 1D lattice gases 
with conserved and 
non-conserved particles density
with initial homogeneous or "shock"-like particle
distributions, $d$-dimensional lattice
gases undergoing continuous exchanges with a reservoir,
as well as an extreme case
of biased tracer
diffusion in a 2D highly-packed
lattice gas containing only 
a single vacancy.
The main emphasis of
our analysis here 
has been put on the question
whether the Einstein relation between
the tracer particle mobility and the diffusivity
is valid, despite the fact that in some cases
the tracer diffusion is anomalously confined
or, in case of regular diffusion, 
the mobility and the diffusivity are non-trivial
functions of the lattice gas density and 
other pertinent parameters. 
For some situations we were able to furnish exact solutions,
which  
show explicitly
that this fundamental  
relation holds. In other instances, to calculate
the tracer particle mobility 
we invoked an approximate approach based
on decoupling of the third-order correlation functions. The tracer
particle diffusivity has been then obtained
by assuming that the Einstein relations holds. 
Extensive numerical 
simulations have been performed, which
confirm our analytical predictions and thus 
the validity of the Einstein relation. 

Apart of it, we have shown that the lattice gas
 particles distribution as seen from 
moving TP 
is strongly inhomogeneous:
the local particle 
density  in front of the TP is higher than the 
average, which means that the
lattice gas particles
tend to accumulate in front of the driven TP, 
creating a sort of a "traffic jam", which impedes
 its motion. The condensed, "traffic jam"-like region in the case when the particles density
is not explicitly 
conserved 
is characterized by a $stationary$ density profile
which vanishes 
as an exponential
function of the distance from the tracer. 
The characteristic
length and the amplitude of the density relaxation function
have been calculated explicitly. 
In one dimensional situations with conserved
particles density the "traffic jam" 
region grows in size in proportion to the TP mean displacement
and no $stationary$ profile exists.
On the other hand, past the TP
the local density is lower than 
the average.  In one dimensional 
situations with conserved
particle density this 
depleted by particles region 
also grows in size
 in proportion to the TP mean 
displacement.
In higher dimensions, 
we have observed that depending on
 whether the number of particles 
is explicitly conserved or not, the local density past the TP
 may tend to the 
average value at large separations from the TP 
in a completely different fashion: In the non-conserved case 
the decay of the density is 
described by an exponential function, 
while for the conserved particles number case it
shows an
$\it algebraic$ dependence on the distance, 
revealing in the latter case
especially strong memory effects and strong 
correlations between the particle distribution in the
lattice gas and the tracer position. 
Further on, we have found
that the terminal velocity $V$  of the TP depends
explicitly on both the 
excess density in the "jammed" region in
front of the TP, as well as on 
the density in the depleted region past 
the tracer. We realized that both densities are 
themselves  dependent on the
 magnitude of the tracer velocity, applied external force,  
as well as on the rate of the adsorption/desorption processes and on the
 rate at which the particles 
can diffuse away of the tracer, which
results in effective non-linear coupling between $V$ and $E$. 
In consequence,    
in the general case (for arbitrary adsorption/desorption 
rates and arbitrary external force), 
$V$ can be found only implicitly, 
as the solution of a transcendental 
equation relating $V$ to the system parameters. 
This equation 
simplifies considerably   
in the limit of a vanishingly small external 
bias; 
in this case we recover  a linear 
force-velocity relation, 
akin to the so-called 
Stokes formula. 
This linear relation 
signifies that  the frictional force exerted on the tracer particle 
by the host medium (the lattice gas) is viscous.
The TP mobility, which is inverse of the Stokes friction coefficient,
thus results from an intricate cooperative behavior.

\end{document}